\documentclass[aps,prl,10pt,twocolumn,groupedaddress,nofootinbib]{revtex4-2}

\usepackage{amsmath,amssymb,bm,graphicx}
\usepackage{xcolor}
\usepackage{pgfplots}
\usepackage{microtype}
\usepackage{hyperref}

\pgfplotsset{compat=1.18}
\hypersetup{hidelinks}
\usepgfplotslibrary{colormaps}

\begin{document}

\title{The Free Scalar as a Thermal Envelope for Three-Dimensional CFTs}

\author{Christopher P. Herzog}
\author{Andreas Stergiou}
\affiliation{Department of Mathematics, King's College London, Strand, London WC2R 2LS, United Kingdom}

\begin{abstract}
    We conjecture that every unitary three-dimensional conformal field theory obeys $c_S/C_T\leq16\pi\zeta(3)$, where $s=c_ST^2$ is the thermal entropy density and $C_T$ the stress-tensor two-point coefficient, with equality for free scalars. We embed this in a stronger conjecture: on $S^1\times H^2$, the $C_T$-normalized hyperbolic heat capacity is bounded above by that of the free scalar for all $T\geq(2\pi R)^{-1}$. Near the conformal temperature, for theories smooth from the high-temperature side and without boundary terms in the modular Hamiltonian, the heat-capacity bound follows from conformal-collider positivity. We test the conjectures in free, interacting large-$N$, and holographic theories. Monte Carlo data for the three-dimensional Ising and $O(N)$ models also obey the flat-space bound.
\end{abstract}

\maketitle

\paragraph{Introduction.} In two spacetime dimensions the leading thermal free energy is fixed by the central charge. In three dimensions this simplification disappears. The sphere free energy $F$, the stress-tensor two-point coefficient $C_T$, and the flat-space thermal entropy coefficient $c_S$ are independent pieces of CFT data \cite{KovtunRitz2008}, all of which can be regarded as measures of the number of degrees of freedom. It is therefore natural to ask whether unitarity and conformal invariance nevertheless impose inequalities among them.

A striking proposal in this direction is the conjecture of Ref.~\cite{BuenoCasini2023} that $C_T/F$ is maximized by the free scalar. Here we propose a complementary inequality involving the thermal coefficient,
\begin{equation}\label{mainbound}
    \boxed{\frac{c_S}{C_T}\leq 16\pi\zeta(3)\,,\qquad s(T)=c_ST^2\,.}
\end{equation}
Although our detailed tests are in three dimensions, the free-scalar upper bound has a natural analogue in $d\geq3$. A perturbative test near four dimensions, including possible infrared and thermal-ordering loopholes, is described in \hyperlink{supplement}{Supplemental Material}. Tests just above two dimensions, where every CFT has the free-scalar value, are also given there.

In $d$ dimensions, write $I_{\mu\nu}(x)=\delta_{\mu\nu}-2x_\mu x_\nu/x^2$. We use the normalization
\begin{equation}
    \begin{gathered}
        \langle T_{\mu\nu}(x)T_{\rho\sigma}(0)\rangle=\frac{C_T}{|x|^{2d}}\,\mathcal I_{\mu\nu,\rho\sigma}(x)\,,\\
        \mathcal I_{\mu\nu,\rho\sigma}=\tfrac12(I_{\mu\rho}I_{\nu\sigma}+I_{\mu\sigma}I_{\nu\rho})-\tfrac1d\delta_{\mu\nu}\delta_{\rho\sigma}\,.
    \end{gathered}
\end{equation}
In $d=3$, one real conformal scalar then has $C_T=3/(32\pi^2)$. Its thermal entropy coefficient is $c_S=3\zeta(3)/(2\pi)$ and hence
\begin{equation}
    \left.\frac{c_S}{C_T}\right|_s=16\pi\zeta(3)=60.4219\ldots\,.
\end{equation}
The free-scalar ratio is unchanged by adding decoupled free scalars. Further, a decoupled topological sector contributes neither to $c_S$ nor to $C_T$.

Several familiar theories lie below this value. A free Dirac fermion has
\begin{equation}
    \frac{c_S}{C_T}=\frac{3}{4}\,16\pi\zeta(3)=45.3164\ldots\,,
\end{equation}
Einstein holography gives (see, e.g., Refs.~\cite{KovtunRitz2008,KlebanovTseytlin1996,GubserKlebanov1997,Anastasiou2022})
\begin{equation}
    \frac{c_S}{C_T}=\frac{4\pi^5}{27}=45.3362\ldots\,,
\end{equation}
and the critical $O(N)$ model at $N=\infty$ gives \cite{Sachdev1993,Petkou1995}
\begin{equation}\label{ONratio}
    \frac{c_S}{C_T}=\frac45\,16\pi\zeta(3)=48.3375\ldots\,.
\end{equation}

Although these examples support the conjecture, a comparison of finitely many CFTs provides limited evidence for a universal bound. In this Letter we propose a stronger inequality that constrains the thermal response on hyperbolic space at every temperature above the conformal temperature. We test this inequality both in individual CFTs and across continuous families of interacting theories.

Before turning to hyperbolic space, where $F$, $C_T$, and $c_S$ can all be embedded in a single thermal function, we note that the thermal coefficient also has a vacuum interpretation. As emphasized by Shaghoulian~\cite{Shaghoulian2015} (see also Ref.~\cite{BobevHongReys2023}), exchanging the thermal and spatial circles relates the leading high-temperature free energy to the vacuum energy on a geometry with one compact spatial circle. In three dimensions,
\begin{equation}\label{casimir}
    E_{\rm vac}(S^1_L\times\mathbb R)=-\frac{\varepsilon_{\rm vac} V}{L^2}\,,\qquad\varepsilon_{\rm vac}=\frac{c_S}{3}\,.
\end{equation}
Here $V$ is the length of the non-compact direction, and fermions are antiperiodic around $S^1_L$, as inherited from the thermal circle. A periodic spatial spin structure defines a different Casimir problem, so the vacuum form of the conjecture refers specifically to this Neveu--Schwarz sector. Thus, the proposed inequality can equally well be written as
\begin{equation}\label{casimirbound}
    \boxed{\frac{\varepsilon_{\rm vac}}{C_T}\leq \frac{16\pi\zeta(3)}{3}\,.}
\end{equation}
Both quantities in Eq.~\eqref{casimirbound} are vacuum data. The conjecture bounds the Casimir energy on a circle in units of $C_T$.

Further thermal-OPE and operator-counting formulations of the bound are given in \hyperlink{supplement}{Supplemental Material}.

\paragraph{A single hyperbolic thermal function.} Place the theory on $S^1_\beta\times H^2_R$ and define
\begin{equation}
    T_0=\frac{1}{2\pi R}\,,\qquad x=\frac{T}{T_0}\,,\qquad\Phi(x)=R^2 s_H(T)\,,
\end{equation}
where $s_H(T)$ is the entropy density on $H^2_R$. The vacuum density matrix reduced to a disk of radius $R$ is conformally equivalent to a thermal density matrix on $\mathbb R_t\times H^2_R$ at $T_0=(2\pi R)^{-1}$ \cite{CasiniHuertaMyers}. In Euclidean signature this state is represented by $S^1_{2\pi R}\times H^2_R$. The same conformal map takes the $q$-sheeted replica geometry, which computes the R\'enyi entropy $S_q$, to $S^1_{2\pi qR}\times H^2_R$. The R\'enyi index is therefore $q=T_0/T=1/x$. We normalize
\begin{equation}\label{Psidef}
    \Psi(x)=\frac{12}{\pi^3 C_T}\Phi(x)\,.
\end{equation}
Three apparently independent CFT quantities are then encoded in elementary properties of one function:
\begin{equation}\label{triangle}
    \boxed{\Psi(1)=\frac{6F}{\pi^4 C_T}\,,\quad\Psi'(1)=1\,,\quad\frac{\Psi(x)}{x^2}\underset{x\to\infty}{\longrightarrow}\frac{3}{\pi^5}\frac{c_S}{C_T}\,.}
\end{equation}
The first relation follows from the equality between the universal disk entanglement entropy and the sphere free energy \cite{CasiniHuertaMyers}. The second is the universal first derivative of spherical R\'enyi entropy and follows from the stress-tensor two-point function \cite{Perlmutter2013}. The third is the high-temperature, or small-R\'enyi-index, limit \cite{BuenoMyersWitczak2015,KusukiOoguriPal2025}, where the curvature radius becomes irrelevant:
\begin{equation}
    R^2s_H(T)\sim c_S R^2T^2=\frac{c_S}{4\pi^2}x^2\,.
\end{equation}

Equation~\eqref{triangle} gives a useful geometric interpretation of the two conjectured CFT bounds. The bound of Ref.~\cite{BuenoCasini2023} on $C_T/F$ says that the free scalar minimizes $\Psi(1)$. Our conjecture says that the free scalar maximizes the coefficient of the large-$x$ growth of $\Psi(x)$. Since every CFT also obeys $\Psi'(1)=1$, the natural quantity to compare is the normalized hyperbolic heat capacity itself.

We therefore conjecture
\begin{equation}\label{heatcapbound}
    \boxed{\Psi'(x)\leq\Psi'_s(x)\,,\qquad x\geq1\,,}\qquad\Psi'(x)=\frac{12R^2}{\pi^3C_T}\frac{c_V(T)}{x}\,,
\end{equation}
where $c_V=T\,d s_H/dT$ and the subscript $s$ in $\Psi'_s(x)$ denotes the free scalar. Thus, at fixed $C_T$ the scalar bounds the hyperbolic heat capacity, with the universal factor $1/x$ dictated by our normalization. (A related free-field envelope of the holographic result was observed numerically for squashed three-spheres after normalization by $C_T$~\cite{BobevBuenoVreys2017}.)

Integrating from $x=1$ gives the weaker entropy envelope
\begin{equation}\label{entropyenvelope}
    \Psi(x)-\Psi(1)\leq\Psi_s(x)-\Psi_s(1)\,,\qquad x\geq1\,.
\end{equation}
Taking the high-temperature limit recovers Eq.~\eqref{mainbound}. This statement is logically independent of the conjectured $C_T/F$ bound because the value at $x=1$ has been subtracted.

The two conjectures nevertheless have a useful combined consequence. For theories to which the $C_T/F$ conjecture applies and for which $\Psi(1)>0$, positivity of the heat capacity gives $\Psi(x)-\Psi(1)\geq0$. Together with $\Psi(1)\geq\Psi_s(1)$ and Eq.~\eqref{entropyenvelope}, this yields
\begin{equation}\label{combinedbound}
    \boxed{\frac{\Psi(x)}{\Psi(1)}\leq\frac{\Psi_s(x)}{\Psi_s(1)}}\quad\Longrightarrow\quad\boxed{\frac{c_S}{F}\leq\left(\frac{c_S}{F}\right)_s=8.9949\ldots\,,}
\end{equation}
where the second inequality is the $x\to\infty$ limit. Thus, the $C_T/F$ bound controls the starting value at $T_0$, while Eq.~\eqref{heatcapbound} controls the subsequent response.

The heat-capacity conjecture is stronger than ordinary thermodynamic stability. Positivity gives $\Psi'(x)\geq0$, but even together with the universal normalization $\Psi'(1)=1$ this does not place an upper bound on the large-$x$ slope. A proof of Eq.~\eqref{heatcapbound} must therefore use constraints specific to conformal field theory. The first such constraint appears already in the second derivative at $x=1$, which is fixed by stress-tensor three-point data.

\paragraph{A local theorem from $\langle TTT\rangle$.} We now show that Eq.~\eqref{heatcapbound} holds near $x=1$ for every CFT whose thermodynamics is smooth above $x=1$ and whose modular Hamiltonian has no boundary term on the entangling surface. Let $h_q$ denote the scaling dimension of the codimension-two twist defect that implements the $q$-fold replica construction across the spherical entangling surface. Its first derivative at $q=1$ is fixed by $C_T$, while its second derivative is determined by an integrated three-point function of the stress tensor \cite{HungMyersSmolkin}. In three dimensions, writing the parity-even $\langle TTT\rangle$ structure in terms of the standard conformal-collider parameter $t_4$ \cite{HofmanMaldacena,Chowdhury2017} (using Eqs.~(2.21), (2.45), (2.46), and (3.14) of Ref.~\cite{HungMyersSmolkin}), one has
\begin{equation}\label{hqexp}
    \frac{h_q}{C_T}=\frac{\pi^3}{24}(q-1)-\frac{\pi^3}{11520}(420+t_4)(q-1)^2+\cdots\,.
\end{equation}
The thermodynamic identity $h_q=\frac q2\int_{1/q}^1\!dx\,x\Phi'(x)$ gives
\begin{equation}
    h'_1=\frac12\Phi'(1)\,,\qquad h''_1=-\frac12\left[\Phi'(1)+\Phi''(1)\right],
\end{equation}
and hence
\begin{equation}\label{psiprimeprime}
    \boxed{\Psi''(1)=\frac34+\frac{t_4}{240}\,.}
\end{equation}

For parity-even theories conformal-collider positivity gives $-4\leq t_4\leq4$. For parity-violating theories the pair $(t_4,\alpha_T)$ lies inside the collider disk \cite{Chowdhury2017}, so in particular $t_4\leq4$. Therefore, every theory without such a boundary term obeys
\begin{equation}\label{localcollider}
    \frac{11}{15}\leq\Psi''(1)\leq\frac{23}{30}\,.
\end{equation}
The upper endpoint is the bosonic parity-even $TTT$ structure, while Einstein gravity has $t_4=0$ and hence $\Psi''(1)=3/4$.

The conformal scalar is exceptional in precisely the way analyzed in Refs.~\cite{HerzogNishioka,LeeEtAl}. The improved scalar stress tensor is associated with a boundary contribution to the modular Hamiltonian,
\begin{equation}\label{boundaryterm}
    H=H_{\rm cov}+2\pi\xi\int_{\partial H^2}\phi^2\,.
\end{equation}
This term does not change the first derivative at $q=1$, because the relevant separated two-point function $\langle T_{\mu\nu}\phi^2\rangle$ vanishes, but it contributes at the next order. In the conventions of Ref.~\cite{HerzogNishioka}, the difference between the full scalar result and the result obtained from the bulk stress tensor alone is
\begin{equation}
    \delta h''_1=-\frac{\pi}{512}\,.
\end{equation}
Using $C_T=3/(32\pi^2)$ gives exactly
\begin{equation}
    \delta\Psi''(1)=\frac12\,.
\end{equation}
Thus,
\begin{equation}\label{scalarcurvature}
    \boxed{\Psi_s''(1)=\frac{19}{15}=\frac{23}{30}+\frac12\,.}
\end{equation}
The excess of $\Psi_s''(1)$ over the conformal-collider upper bound is thus accounted for by the scalar modular-Hamiltonian boundary term. What is new here is its role in separating the free-scalar thermal envelope from the ordinary conformal-collider bound.

Combining Eqs.~\eqref{localcollider} and \eqref{scalarcurvature} gives
\begin{equation}\label{localheatcap}
    \boxed{\Psi_s'(x)-\Psi'(x)\geq\frac12(x-1)+O((x-1)^2)\,,}
\end{equation}
provided the hyperbolic thermodynamics admits a right-sided Taylor expansion through this order at $x=1$. Thus, the differential conjecture holds sufficiently close to $x=1$, with a finite margin; integration gives an entropy gap $\frac14(x-1)^2+O((x-1)^3)$. The remaining qualification is a boundary term built from a symmetry-allowed dimension-one scalar, as $\phi^2$ is for the free scalar. Such an operator is marginal on the entangling surface. Theories with such a term require separate analysis and are not covered by Eq.~\eqref{localheatcap}.

The restriction to $x\geq1$ is essential: for theories smooth across $x=1$, the linear term in Eq.~\eqref{localheatcap} reverses sign just below $x=1$. In the free Dirac and Einstein examples the reversed ordering persists to $x=0$; the conventional critical $O(N)$ saddle has no continuation to $x<1$.

\paragraph{Global tests.} We first test Eq.~\eqref{heatcapbound} in theories whose hyperbolic response can be computed. We then test the flat-space bound, Eq.~\eqref{mainbound}, in theories whose hyperbolic response is unknown.

For free fields, the calculation follows by differentiating the thermal free energies and using the $H^2$ spectral densities of Ref.~\cite{KlebanovPufuSachdevSafdi}. For a one-particle mode of energy $E$, let $z=\beta E$ be the dimensionless thermal energy. The bosonic and fermionic single-mode entropies are
\begin{align}
    s_B(z)&=\frac{z}{e^z-1}-\log(1-e^{-z})\,,\\
    s_F(z)&=\log(1+e^{-z})+\frac{z}{e^z+1}\,.
\end{align}
For later use, define the dimensionless effective mass $m$ by $E_p=R^{-1}\sqrt{p^2+m^2}$, and write the bosonic entropy as
\begin{equation}\label{scalarPhi}
    \Phi_B(x;m)=\int_0^\infty\frac{dp}{2\pi}\,p\tanh(\pi p)\,s_B\!\left(\frac{2\pi\sqrt{p^2+m^2}}{x}\right),
\end{equation}
where $p$ is the dimensionless principal-series momentum on $H^2$. For the free conformal scalar $m=0$, so $\Phi_s(x)=\Phi_B(x;0)$; for the interacting models below, $m$ is fixed by a gap equation. The free-Dirac entropy is instead
\begin{equation}
    \Phi_F(x)=\int_0^\infty\frac{dp}{\pi}\,p\coth(\pi p)\,s_F(2\pi p/x)\,.
\end{equation}
At high temperature, the free scalar exhibits a logarithmic correction,
\begin{equation}\label{scalarhighT}
    \Psi_s(x)=\frac{48\zeta(3)}{\pi^4}x^2-\frac{8}{3\pi^2}\log x+O(1)\,.
\end{equation}
This logarithm reflects the absence of a spectral gap. The conformal coupling $\xi{\cal R}=-1/(4R^2)$ cancels the gap $1/(4R^2)$ of the $H^2$ Laplacian, so the scalar spectrum extends down to $E=0$. Without interactions, no thermal mass screens these low-energy modes. The thermal effective action on $S^1\times H^2$ of Ref.~\cite{ParmentierBobev2026} gives the same coefficient.

\begin{figure}[t]
    \includegraphics{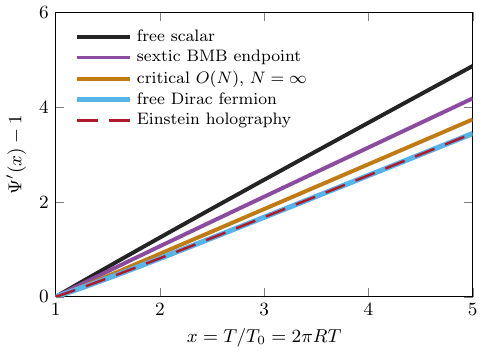}
    \caption{Normalized hyperbolic heat-capacity response $\Psi'(x)-1$. All curves start at zero because $\Psi'(1)=1$. The free scalar is an upper envelope for the critical $O(N)$ model, the BMB endpoint of the stable sextic line, a free Dirac fermion, and Einstein holography. The dashed Einstein curve nearly coincides with the Dirac curve.}
    \label{fig:envelope}
\end{figure}

Differentiating the corresponding Bose and Fermi entropy integrals gives the exact heat-capacity curves shown in Fig.~\ref{fig:envelope}. The scalar lies strictly above the fermion for all $x>1$; this ordering also follows analytically from the spectral integrals (\hyperlink{supplement}{Supplemental Material}).

Einstein holography supplies a very different example. The thermal state on $H^2$ is represented by the hyperbolic Schwarzschild-AdS black hole \cite{CasiniHuertaMyers,HungMyersSmolkinYale}. Let $r_h$ denote the horizon radius and $\widetilde L$ the AdS curvature radius. Defining $y=r_h/\widetilde L$, its temperature satisfies
\begin{equation}
    x=\frac{3y^2-1}{2y}\,,\qquad y\geq 1\,,
\end{equation}
and, after the normalization in Eq.~\eqref{Psidef},
\begin{equation}
    \Psi_{\rm Ein}(x)=y(x)^2\,.
\end{equation}
Here $y(x)=(x+\sqrt{x^2+3})/3$, and the solution exists for all $x\geq0$. The Einstein curve obeys $\Psi_{\rm Ein}''\leq8/9$, whereas $\Psi_s''>96\zeta(3)/\pi^4\approx1.18$ for $x\geq1$. Integrating from $x=1$ proves Eq.~\eqref{heatcapbound} for all $x>1$ (\hyperlink{supplement}{Supplemental Material}).

A considerably sharper test is provided by the critical $O(N)$ model at $N=\infty$. At leading order its sphere free energy and $C_T$ are those of $N$ free scalars, and its parity-even stress-tensor three-point function is at the bosonic collider endpoint $t_4=4$. Thus, on the high-temperature side, its entropy function $\Psi(x)$ shares the free scalar's value and first derivative at $x=1$, while its curvature $\Psi''(1^+)=23/30$ saturates the collider bound \eqref{localcollider} without the free scalar's additional $\phi^2$ boundary contribution. The boundary contribution is absent because the interaction replaces $\phi^2$ by the Hubbard--Stratonovich field $\sigma$, whose dimension is two.

On $S^1\times H^2$ the conventional saddle generates a dimensionless thermal mass $m(x)$, determined by the large-$N$ gap equation (Eq.~(3.9) of Ref.~\cite{Mukherjee2015})
\begin{equation}\label{ONgap}
    0=\int_0^\infty dp\left[\frac{p\tanh(\pi p)}{\sqrt{p^2+m^2}}\coth\!\left(\frac{\pi\sqrt{p^2+m^2}}{x}\right)-1\right].
\end{equation}
For $x\geq1$ this saddle is well-behaved, with $m(1)=0$. At leading order, the entropy is $\Phi(x)=N\Phi_B(x;m(x))$, using Eq.~\eqref{scalarPhi}. The saddle reproduces this curvature from the $x>1$ side, even though the thermodynamics is not analytic across $x=1$. On this conventional saddle $m(x)$ increases with $x$, and the single-mode heat capacity decreases with mass. These facts establish Eq.~\eqref{heatcapbound} analytically for all $x>1$ at leading $N$ (\hyperlink{supplement}{Supplemental Material}). At high temperature, $2\pi m(x)/x\to2\log\varphi$, where $\varphi=(1+\sqrt5)/2$, reproducing the flat-space thermal mass \cite{Sachdev1993}, and the entropy coefficient tends to Eq.~\eqref{ONratio}. The conventional saddle has no real solution for $x<1$ \cite{Mukherjee2015}; we use the $x\geq1$ branch.

The large-$N$ sextic line gives a particularly direct test because it contains the free scalar itself. For the interaction $g_6(\phi\cdot\phi)^3/(6N^2)$, define $\alpha=\sqrt{g_6}/(4\pi)$. The stable large-$N$ conformal line is $0\leq\alpha\leq1$, ending at the Bardeen--Moshe--Bander point \cite{BMB,Goykhman2021,BMBmodern}. On the conformal vacuum saddle, $F$ and $C_T$ retain their free-field leading values. Write ${\cal I}(m,x)$ for the integral in Eq.~\eqref{ONgap}. The sextic saddle then obeys~\cite{Mukherjee2015,Goykhman2021}
\begin{equation}\label{sexticgap}
    m=\alpha\,{\cal I}(m,x)\,.
\end{equation}
The same monotone-mass argument establishes Eq.~\eqref{heatcapbound} on the conventional saddle throughout this line for $x>1$ (\hyperlink{supplement}{Supplemental Material}). At fixed $x>1$ the response decreases with the sextic coupling; the BMB endpoint is shown in Fig.~\ref{fig:envelope}.

At the BMB endpoint, $\alpha=1$, the flat-space limit gives $m=\frac{x}{2\pi}\log 2$, and hence
\begin{equation}\label{BMBratio}
    \left.\frac{c_S}{C_T}\right|_{\rm BMB}=14\pi\zeta(3)=52.8692\ldots\,.
\end{equation}
Thus, the stable sextic line runs from free-scalar saturation down to $7/8$ of its thermal coefficient, rather than to the critical-$O(N)$ value. Its right-sided curvature at the conformal temperature is $\Psi_{\rm BMB}''(1^+)=19/15-\pi^2/[2(2+\pi)^2]=1.0799967\ldots$ (see \hyperlink{supplement}{Supplemental Material}), below the free scalar but above the ordinary collider upper bound $23/30$. This illustrates the separate behavior of theories with a dimension-one $\phi^2$ operator, not covered by Eq.~\eqref{localheatcap}. At finite $N$ the fixed-line interpretation is more subtle, so we use this example only as a strict large-$N$ test.

In every example with a known hyperbolic response we also find the stronger ordering $\Psi''(x)\leq\Psi_s''(x)$ for $x>1$, which integrates to Eq.~\eqref{heatcapbound}. Unlike positivity of the heat capacity, no general CFT principle is known to imply it.

The remaining tests concern only the flat-space bound, because the hyperbolic response is not known. In holography, homogeneous planar Einstein-scalar black branes obey the stronger bound $c_S/C_T\leq4\pi^5/27$, even with source-free scalar hair. This follows from warp-factor monotonicity (\hyperlink{supplement}{Supplemental Material}).

Large-$N$ Chern--Simons (CS) vector models provide further tests. Write $\lambda=N/k$ for the 't Hooft coupling, where $k$ is the renormalized CS level, and restrict by parity to $0\leq\lambda<1$. At planar order the flat-space bound holds analytically for CS theories with a single fundamental scalar or fermion, at either the regular or the critical fixed point (\hyperlink{supplement}{Supplemental Material}). The thermal free energies follow from the planar saddle of Ref.~\cite{AharonyThermal}, and the stress-tensor normalization is
\begin{equation}\label{CSCT}
    C_T(\lambda)=C_T(0)\frac{\sin(\pi\lambda)}{\pi\lambda}\,.
\end{equation}
The regular-boson (RB) theory also has a marginal sextic coupling $\lambda_6(\phi^\dagger\phi)^3/(6N^2)$ at $N=\infty$. Varying $\lambda_6$ leaves the planar $\langle TT\rangle$ and separated-point $\langle TTT\rangle$ unchanged~\cite{AharonyCorrelators}, but changes $c_S$. For example, two stable couplings at $\lambda=1/2$ give
\begin{equation}\label{sameTTT}
    \left.\frac{c_S}{C_T}\right|_{\lambda=1/2}=\begin{cases}
        49.0338\,,&\lambda_6=6\pi^2\,,\\
        48.8231\,,&\lambda_6=12\pi^2\,.
    \end{cases}
\end{equation}
The stability conditions and the full $(\lambda,\lambda_6)$ dependence are given in the \hyperlink{supplement}{Supplemental Material}, using Ref.~\cite{Dey2018}.

Finite-$N$ interacting CFTs can also be tested using critical Casimir amplitudes. For a three-dimensional classical critical system on a periodic slab, the excess free energy per transverse area is $\Delta/L^2$. This is the Casimir energy of Eq.~\eqref{casimir} with $\Delta=-\varepsilon_{\rm vac}$, so
\begin{equation}\label{CasimirToCS}
    c_S=-3\Delta\,.
\end{equation}
Periodic-slab Monte Carlo amplitudes for the Ising and $O(N)$ models, combined with bootstrap determinations of $C_T$, give the endpoint ratios in Table~\ref{tab:evidence}. These are finite-$N$, strongly interacting, nonholographic tests of Eq.~\eqref{mainbound}. For $O(4)$ and $O(6)$, $C_T$ is taken at its bootstrap lower bound~\cite{KosPolandSimmons2014}, so these two ratios are upper estimates.

Published thermal and stress-tensor $1/N$ expansions \cite{Petkou1995,ChubukovSachdevYe1994,DiatlykPopovWang2024,DiabEtAl2016,GiombiTarnopolskyKlebanov2016} give $(c_S/C_T)_{O(N)}=48.3376+1.5557/N+O(N^{-2})$; the Gross--Neveu and conformal QED$_3$ endpoints also satisfy Eq.~\eqref{mainbound} in their controlled large-$N_f$ regimes (\hyperlink{supplement}{Supplemental Material}).

Table~\ref{tab:evidence} summarizes the tests.
\begin{table}[b]
    \caption{Thermal ratios and status of the heat-capacity test. ``Full'' refers to the $x>1$ response on the indicated branch; ``endpoint'' tests only the flat-space limit.}
    \label{tab:evidence}
    \begin{ruledtabular}
        \begin{tabular}{lcc}
            Theory & $c_S/C_T$ & Test \\
            \hline
            free scalar & $60.422$ & saturation\\
            critical $O(N)$, $N=\infty$ & $48.338$ & full\\
            sextic line, $N=\infty$ & $52.869$--$60.422$ & full\\
            Einstein holography & $45.336$ & full\\
            free Dirac & $45.316$ & full\\
            3d Ising~\cite{BulgarelliCaselle2025,ChangIsing2025} & $51.08(13)$ & endpoint\\
            $O(2)$~\cite{BulgarelliCaselle2025,ReehorstTrevisaniVichi} & $50.84(27)$ & endpoint\\
            $O(3)$~\cite{BulgarelliCaselle2025,Chester2021} & $50.32(23)$ & endpoint\\
            $O(4)$~\cite{KosPolandSimmons2014,BulgarelliCaselle2025} & $50.05(24)$ & endpoint\\
            $O(6)$~\cite{KosPolandSimmons2014,BulgarelliCaselle2025} & $49.66(29)$ & endpoint\\
            RB CS-matter & $45.316$--$60.422$ & endpoint\\
            other CS-matter & $<60.422$ & endpoint\\
            Gross--Neveu & $45.32+0.08/N_f$ & endpoint\\
            QED$_3$ & $45.32+0.90/N_f$ & endpoint
        \end{tabular}
    \end{ruledtabular}
\end{table}

\paragraph{Discussion.} The free scalar's behavior near $x=1$ and as $x\to\infty$ provides a partial explanation of its conjectured role in setting the upper bound on $c_S/C_T$. Near $x=1$, its $\phi^2$ boundary term in the modular Hamiltonian lifts the thermal curvature above the ordinary collider bound. At large $x$, the conformal coupling cancels the $H^2$ spectral gap, leaving a gapless bosonic spectrum and the logarithm in Eq.~\eqref{scalarhighT}. Interactions generate a thermal mass in the large-$N$ bosonic examples, removing the latter effect.

The CS examples in Eq.~\eqref{sameTTT} show why the local collider argument, Eq.~\eqref{localcollider}, cannot determine the global bound: two theories with identical leading $C_T$ and separated-point $\langle TTT\rangle$ can have different thermal coefficients. Thus, control of the full thermal response requires information beyond those local data.

Thermodynamic positivity gives $\Psi'(x)\geq0$ because heat capacity is a Hamiltonian variance, but does not bound this positive quantity from above in units of $C_T$. The cycle-exchange interpretation in Eq.~\eqref{casimirbound} suggests investigating whether reflection positivity, modular covariance, or a vacuum variational principle on $S^1\times\mathbb R$ can constrain the normalized Casimir amplitude.

The stronger pointwise conjecture is vulnerable to changes of thermal phase: a first-order transition at $x>1$ with positive latent heat would produce a positive delta function in $\Psi'(x)$ and violate Eq.~\eqref{heatcapbound}, even if the flat-space bound survives. To our knowledge, no such transition has been established for a unitary three-dimensional CFT on $S^1\times H^2$ with vacuum R\'enyi boundary conditions. Further tests could include a first-principles CS-matter calculation on this space.

The flat-space bound may extend to $d\geq3$. Near four dimensions, every real multiscalar fixed point with a stable symmetric thermal state obeys it at leading $O(\varepsilon)$ (\hyperlink{supplement}{Supplemental Material}). Static Matsubara modes give an opposing $O(\varepsilon^{3/2})$ correction, which is subleading at sufficiently small $\varepsilon$. More seriously, persistent thermal symmetry breaking can contribute with the opposite sign already at $O(\varepsilon)$~\cite{Chai2020}. At the ordered $O(1)\times O(n)$ biconical fixed points, this contribution cancels only part of the $O(\varepsilon)$ reduction, so the bound still holds at this order for every finite $n$ (\hyperlink{supplement}{Supplemental Material}). The cancellation becomes complete as $n\to\infty$: an emergent flat direction leaves the thermal coefficient equal to that of free scalars at leading order, though the CFT remains interacting~\cite{Chai2020,SmolkinYung2026}. No general inequality for thermally ordered phases, or higher-dimensional hyperbolic envelope, has been established.

In $d=2$, $c_S$ and $C_T$ are both proportional to the central charge, so every CFT has the free-scalar value of $c_S/C_T$. Just above two dimensions, the Gross--Neveu fixed point lies below the free-scalar value at the same $d$. A dimensional-reduction argument predicts a ratio $1-\frac32\varepsilon+\ldots$ to the free-scalar value in $d=2+\varepsilon$ for sigma models with compact positive-Einstein targets. The formal continuation of Einstein holography lies above the free-scalar value for 
$2<d<2.147\ldots$ (\hyperlink{supplement}{Supplemental Material}).

\begin{acknowledgments}
    This work was supported by STFC grants ST/X000753/1 and UKRI4231. AS is supported by the Royal Society under grant URF{\textbackslash}R1{\textbackslash}211417. OpenAI ChatGPT (GPT-5.6, 6.1 Sol and 6 Astra), Anthropic Claude (Opus 5.5, Fable 5.1) assisted with literature synthesis, proof attempts, cross-checks, code, figures, and revision. The authors directed and checked the analysis and take responsibility for the manuscript.
\end{acknowledgments}

\paragraph*{Data availability.} Published Monte Carlo and bootstrap inputs are cited. Numerical data and code for the calculations performed here are available from the authors.

\bibliography{references}

\clearpage
\onecolumngrid
\begin{center}
    {\large\bfseries\hypertarget{supplement}{Supplemental Material}}
\end{center}
\setcounter{equation}{0}
\setcounter{figure}{0}
\setcounter{table}{0}
\renewcommand{\theequation}{S\arabic{equation}}
\renewcommand{\thefigure}{S\arabic{figure}}
\renewcommand{\thetable}{S\arabic{table}}
\renewcommand{\theHequation}{S\arabic{equation}}
\renewcommand{\theHfigure}{S\arabic{figure}}
\renewcommand{\theHtable}{S\arabic{table}}

\section*{Methods}

The idea for this Letter emerged from the following prompt that Herzog issued to GPT-5.6 Sol and an analogous one to Claude Opus 5.5:
\begin{quote}
I would like you to write for me a short article that could be published in Physical Review Letters.  I would like you to choose the question, do the analysis and write up the results. The constraints I put on the question are 1) that it should be in the hep-th set of topics (so that I have a hope of evaluating the arguments and results); 2) the answer can be written up in a short format of 5 pages or less; 3) the result is of significance enough that it can be published in PRL.  I want to break up the task into several steps.  The first is to find a good question, so give me three candidates.  I'm also going to involve Claude in the process.  Claude is going to find three candidates too, and the next step will be to evaluate each others' candidate questions.
\end{quote}
The resulting six ideas were then rank ordered by the agents and consensus was reached.  Ideas one and three were tossed out after initial explorations, and the second idea [main text, (1)], proposed by Claude, was used for the letter.  After ChatGPT pointed out the relevance of the sphere free energy conjecture \cite{BuenoCasini2023}, Claude proposed the extended conjecture on $S^1 \times H^2$.  

All of the calculations were carried out by agents, and the initial draft of the manuscript was also prepared by agents, within a few hours. The resulting prose was often insufficiently detailed, and key references were often missing. GPT-6 Astra and Claude Fable 5.1 were recruited to check the examples, generate a few further examples and try, unsuccessfully, to find a proof of the conjecture. After two weeks of work, we verified all of the computations by hand or using Mathematica, clarified many of the explanations, and suggested a few key additional details.  For example, we suggested the inclusion of~\cite{HerzogNishioka}, the conjecture in terms of CFT data, Eqs.~\eqref{cSspectrum}--\eqref{countingbound}, and the section on multiscalar theories.  

Despite the essential use of AI in all aspects of manuscript preparation, we feel that the result is reliable and of sufficient interest to be published.

\section*{Thermal OPE and spectral reformulations}

For a scalar primary $\mathcal O$ of dimension $\Delta$, its thermal two-point function on $S^1_\beta\times\mathbb R^2$ admits an OPE in which the stress tensor appears with a coefficient conventionally denoted $a_T$. The Ward identity fixes the zero-temperature $\mathcal O\mathcal O T$ coupling, while the finite-temperature one-point function of $T_{\mu\nu}$ is determined by the thermal pressure. In the conventions of Ref.~\cite{Iliesiu2018}, in $d=3$ one obtains
\begin{equation}\label{aTrelation}
    a_T=\frac{\Delta}{4\pi}\frac{c_S}{C_T}\,.
\end{equation}
This relation goes back to Cardy~\cite{Cardy1987} and was applied to the large-$N$ $O(N)$ model in Ref.~\cite{PetkouVlachos1999}. The conjectured bound $c_S/C_T\leq16\pi\zeta(3)$ is therefore equivalent to
\begin{equation}
    a_T\leq 4\zeta(3)\Delta\,.
\end{equation}
For the free scalar with $\Delta=1/2$, the image sum on the thermal circle gives coefficients proportional to $\zeta(J+1)$; the spin-two term reproduces $a_T=2\zeta(3)$ and saturates this bound. This is not a proof: generic thermal OPE coefficients are not sign definite, and the discontinuity entering the thermal Lorentzian inversion formula~\cite{Iliesiu2018} does not provide an obvious positive measure. In the large-$N$ $O(N)$ and Gross--Neveu models, applying the inversion formula to the thermal two-point function yields the gap equation for the thermal mass~\cite{PetkouStergiou2018}.

The thermal coefficient is itself spectral CFT data. Place the theory on $S^1_\beta\times S^2$, with $S^2$ of unit radius. Its finite-temperature one-point functions are Boltzmann-weighted sums of diagonal matrix elements over local operators~\cite{ElShowkPapadodimas,BenjaminLeeOoguriSimmonsDuffin}. For the integrated stress tensor, the Ward identity relates this one-point function to the Hamiltonian; only the energy spectrum survives. In the high-temperature limit,
\begin{equation}\label{cSspectrum}
    c_S=\frac{3}{4\pi}\lim_{\beta\to0}\beta^2\log\sum_{\mathcal O}e^{-\beta\Delta_{\mathcal O}}\,,
\end{equation}
where the sum includes all local operators, descendants, and spin components, with multiplicity. Thus, the flat-space conjecture can be written as
\begin{equation}\label{spectralbound}
    \lim_{\beta\to0}\beta^2\log\sum_{\mathcal O}e^{-\beta\Delta_{\mathcal O}}\leq\frac{64\pi^2\zeta(3)}{3}\,C_T=2\zeta(3)\frac{C_T}{C_{T,s}}\,.
\end{equation}
Descendants multiply the contribution of a primary by at most $(1-e^{-\beta})^{-3}$, changing the logarithm only by $O(\log\beta)$. Accordingly, the leading bound is unchanged if the sum is restricted to primaries, weighted by their spin degeneracy $2J+1$.

Let $N(E)$ count local operators, with multiplicity, of scaling dimension at most $E$. Positivity gives $Z(\beta)\geq N(E)e^{-\beta E}$, and hence
\begin{equation}
    \log N(E)\leq\inf_{\beta>0}\!\bigl[\beta E+\log Z(\beta)\bigr]\,,\qquad Z(\beta)=\sum_{\mathcal O}e^{-\beta\Delta_{\mathcal O}}\,.
\end{equation}
If $\log Z(\beta)\sim A/\beta^2$, the minimizing $\beta$ is asymptotically $(2A/E)^{1/3}$, giving a coefficient $\frac32(2A)^{1/3}$ for $E^{2/3}$. Conversely, the Laplace representation $Z(\beta)=\beta\int_0^\infty e^{-\beta E}N(E)\,dE$ recovers the corresponding upper bound on $A$ by a saddle-point estimate. Equation~\eqref{spectralbound} is consequently equivalent, at the level of upper asymptotic growth bounds, to
\begin{equation}\label{countingbound}
    \limsup_{E\to\infty}\frac{\log N(E)}{E^{2/3}}\leq\frac32\left(4\zeta(3)\frac{C_T}{C_{T,s}}\right)^{1/3}\,.
\end{equation}
Thus, no unitary CFT is conjectured to have more rapid asymptotic growth of local operators at fixed $C_T$ than the free scalar. In two dimensions, modular invariance fixes the analogous growth in terms of the central charge. No corresponding modular constraint on $S^1\times S^2$ is presently known.

\section*{Planar Einstein gravity with scalar hair}

Consider four-dimensional Einstein gravity with scalar fields $\phi^a$ and positive-definite sigma-model metric $G_{ab}$, normalized by
\begin{equation}
    S=\frac{1}{16\pi G}\int d^4x\,\sqrt{-g}\left[R-\frac12G_{ab}(\phi)\,\partial_\mu\phi^a\partial^\mu\phi^b-V(\phi)\right].
\end{equation}
Assume a source-free asymptotically AdS vacuum of radius $L$. For a homogeneous planar black brane write $ds^2=e^{2A(r)}[-h(r)dt^2+d\vec x^{\,2}]+dr^2/h(r)$, with $\phi^a=\phi^a(r)$ and primes denoting $r$ derivatives. The radial Einstein equations imply
\begin{equation}
    A''=-\frac14G_{ab}\phi^{a\prime}\phi^{b\prime}\leq0\,,\qquad h''+3A'h'=0\,.
\end{equation}
The first relation is the four-dimensional scalar specialization of the warp-factor concavity in Eqs.~(4.5) and (4.8), (4.9) of Freedman \emph{et al.}~\cite{Freedman1999}; the second and the finite-temperature generalization of the first are analogous to Eqs.~(25a), (25b) of Gubser and Nellore~\cite{GubserNellore2008}, which use five bulk dimensions and a different radial gauge. In particular, $(e^{3A}h')'=0$. With $A'(\infty)=1/L$, the former implies $A(r)\geq A_h+(r-r_h)/L$. If $J=\int_{r_h}^\infty e^{-3A(r)}dr$, then $J\leq Le^{-3A_h}/3$, and
\begin{equation}
    T=\frac{e^{-2A_h}}{4\pi J}\geq\frac{3e^{A_h}}{4\pi L}\,,\qquad s=\frac{e^{2A_h}}{4G}\,.
\end{equation}
Consequently $c_S=s/T^2\leq4\pi^2L^2/(9G)$. Using the Einstein value $C_T=3L^2/(\pi^3G)$ yields
\begin{equation}
    \frac{c_S}{C_T}\leq\frac{4\pi^5}{27}\,.
\end{equation}
Equality requires vanishing scalar gradients. This concerns only regular, isotropic, source-free planar solutions and the flat-space endpoint, not hyperbolic-horizon transitions.

\paragraph{Continuation away from three dimensions.} For a planar Einstein black brane, the ratio of the thermal entropy coefficient to the stress-tensor central charge can be extracted from Eq.~(5) of Kovtun and Ritz~\cite{KovtunRitz2008}. In our conventions,
\begin{equation}
    \left.\frac{c_S}{C_T}\right|_{\rm Ein}=\left(\frac{4\pi}{d}\right)^d\frac{(d-1)\pi^{d/2}\Gamma(d/2)}{2(d+1)\Gamma(d)}\,.
\end{equation}
Dividing by the free-scalar value in Eq.~\eqref{Sscalarbenchmark} below gives
\begin{equation}\label{SEinsteinRatioD}
    R_{\rm Ein}(d)\equiv\frac{(c_S/C_T)_{\rm Ein}}{(c_S/C_T)_s}
    =\frac{\pi}{2}\left(\frac{4\pi}{d}\right)^{d-1}
    \frac{\Gamma(d/2)^2}{(d+1)\Gamma(d+1)\zeta(d)}\,.
\end{equation}
At $d=2$ the ratio is one. Expanding about this point gives
\begin{equation}
    R_{\rm Ein}(2+\varepsilon)
    =1+\left[\log(2\pi)-\frac{7}{3}-\frac{\zeta'(2)}{\zeta(2)}\right]\varepsilon+O(\varepsilon^2)
    =1+0.074505\,\varepsilon+O(\varepsilon^2)\,.
\end{equation}
For $d\geq2$, the formally continued Einstein ratio exceeds one only for $2<d<d_\times$, equals one at both endpoints, and lies below one for all $d>d_\times$, where $d_\times=2.147469\ldots$. The comparison at noninteger $d$ concerns the analytic continuation of the black-brane formula. The existence of a unitary boundary CFT realizing that continuation is a separate question.

\section*{Analytic hyperbolic envelope for free and large-\texorpdfstring{$N$}{N} fields}

We give analytic arguments for the full-curve comparisons in the Letter, first for free fields and Einstein holography, and then for interacting scalar saddles.

\paragraph{Free Dirac versus free scalar.} Set $w_B(z)=z^3e^z/(e^z-1)^2$ and $w_F(z)=z^3e^z/(e^z+1)^2$. Differentiating the free-field spectral integrals gives
\begin{equation}
    \Psi_s'(x)=\frac{16x}{\pi^4}\int_0^\infty dz\,w_B(z)\tanh\!\frac{xz}{2}\,,\quad\Psi_F'(x)=\frac{16x}{\pi^4}\int_0^\infty dz\,w_F(z)\coth\!\frac{xz}{2}\,.
\end{equation}
Writing $t=\tanh(xz/2)$ and $t_1=\tanh(z/2)$, one finds
\begin{equation}
    \Psi_s'(x)-\Psi_F'(x)=\frac{4x}{\pi^4}\int_0^\infty dz\,\frac{z^3(t^2-t_1^2)}{\sinh^2(z/2)\,t}>0\qquad(x>1)\,,
\end{equation}
using only that $\tanh$ is increasing.

\paragraph{Einstein holography versus free scalar.} Let $a_s=96\zeta(3)/\pi^4=1.1846\ldots$ be the asymptotic slope of $\Psi_s'$. Since $\int_0^\infty w_B\,dz=6\zeta(3)$, writing $\tanh(xz/2)=1-2/(e^{xz}+1)$ and substituting $v=xz$ gives
\begin{equation}
    \Psi_s'(x)=a_sx-\frac{32}{\pi^4}\int_0^\infty dv\,\frac{w_B(v/x)}{e^v+1}\,.
\end{equation}
Differentiating and returning to $z=v/x$,
\begin{equation}
    \Psi_s''(x)=a_s+\frac{32}{\pi^4}G(x)\,,\qquad G(x)=\int_0^\infty dz\,\frac{z\,w_B'(z)}{e^{xz}+1}\,.
\end{equation}
Since $(\log w_B)'=3/z-\coth(z/2)$ decreases strictly, $w_B$ has a single maximum at some $z_*$. So $z\,w_B'$ is positive for $z<z_*$ and negative for $z>z_*$. For $x\geq1$ the ratio $r(z)=(e^z+1)/(e^{xz}+1)$ is positive and non-increasing, because $e^y/(e^y+1)$ increases with $y$. Splitting the integral at $z_*$ therefore gives $G(x)\geq r(z_*)\,G(1)$. Direct evaluation gives $G(1)=19\pi^4/480-3\zeta(3)=0.2496\ldots$, which is equivalent to $\Psi_s''(1)=19/15$. Hence $\Psi_s''(x)>a_s$ for $x\geq1$, and integrating from $\Psi_s'(1)=1$ gives
\begin{equation}
    \Psi_s'(x)\geq1+a_s(x-1)\,,\qquad x\geq1\,.
\end{equation}
For Einstein holography $\Psi_{\rm Ein}=y^2$ with $y=(x+\sqrt{x^2+3})/3$. Setting $x=\sqrt3\sinh\theta$ and $v=e^{-2\theta}$ gives
\begin{equation}
    \Psi_{\rm Ein}''(x)=\frac89\,\frac{1+3v}{(1+v)^3}\leq\frac89\,,
\end{equation}
because $(1+v)^3\geq1+3v$. Integrating from $\Psi_{\rm Ein}'(1)=1$ gives $\Psi_{\rm Ein}'(x)\leq1+\frac89(x-1)$. Since $a_s>8/9$,
\begin{equation}
    \Psi_s'(x)-\Psi_{\rm Ein}'(x)\geq\left(a_s-\frac89\right)(x-1)>0\,,\qquad x>1\,.
\end{equation}
This assumes that the hyperbolic black hole without scalar hair is the dominant saddle for $x\geq1$.

\paragraph{Interacting scalar saddles and monotone thermal mass.} For the critical $O(N)$ model and the large-$N$ sextic scalar line we assume that the homogeneous, vacuum-connected saddle gives the leading-$N$ thermodynamics on $S^1\times H^2$ for $x\geq1$, with the standard replica boundary conditions. The following argument establishes the envelope on those saddles, not the exclusion of competing phases.

Write $\omega=\sqrt{p^2+m^2}$. The subtracted tadpole integral in the Letter's critical-$O(N)$ gap equation is
\begin{equation}\label{SintI}
    \mathcal I(m,x)=\int_0^\infty dp\,\left[\frac{p\tanh(\pi p)}{\omega}\coth\!\left(\frac{\pi\omega}{x}\right)-1\right].
\end{equation}
The critical model obeys $\mathcal I(m,x)=0$, whereas on the sextic line $m=\alpha\mathcal I(m,x)$ with $0\leq\alpha\leq1$. For $m>0$, differentiation at fixed $m$ or $x$, respectively, gives
\begin{align}
    \partial_x\mathcal I&=\frac{\pi}{x^2}\int_0^\infty dp\,p\tanh(\pi p)\,\operatorname{csch}^2\!\left(\frac{\pi\omega}{x}\right)>0\,,\\
    \partial_m\mathcal I&=-m\int_0^\infty dp\,p\tanh(\pi p)\left[\frac{\coth(\pi\omega/x)}{\omega^3}+\frac{\pi}{x\omega^2}\operatorname{csch}^2\!\left(\frac{\pi\omega}{x}\right)\right]<0\,.\label{SIderivatives}
\end{align}
Thus, $\mathcal I$ increases with $x$ and decreases with $m$; the latter monotonicity also holds at $m=0$ by continuity. At $x=1$, $\mathcal I(0,1)=0$ because $\tanh(\pi p)\coth(\pi p)=1$. For $x>1$, $\mathcal I(0,x)>0$, since the hyperbolic cotangent is strictly decreasing in its argument. Finally, at fixed $x$, $\mathcal I(m,x)=-m+O(m^{-1})$ as $m\to\infty$. Indeed, separating $\tanh(\pi p)=1+[\tanh(\pi p)-1]$ and $\coth u=1+2/(e^{2u}-1)$, the first term integrates to $\int_0^\infty(p/\omega-1)\,dp=-m$; the curvature correction is $-1/(24m)+O(m^{-3})$, while the thermal correction is exponentially small at fixed $x$. The critical gap equation therefore has a unique root $m(x)>0$; the sextic equation has a unique root for each $\alpha>0$, since $m-\alpha\mathcal I(m,x)$ increases strictly from a negative value to a positive one. Implicit differentiation now establishes
\begin{equation}\label{Smassmonotone}
    m'_{\rm crit}(x)=-\frac{\partial_x\mathcal I}{\partial_m\mathcal I}>0\,,\qquad m'_\alpha(x)=\frac{\alpha\partial_x\mathcal I}{1-\alpha\partial_m\mathcal I}>0\,,\qquad (x>1)\,.
\end{equation}
The $m\to0^+$ derivative requires care: the integral is not uniformly differentiable at $p=m=0$. Its one-sided expansion near $(m,x)=(0,1)$ is given below in the sextic-curvature section.

For $z=2\pi\sqrt{p^2+m^2}/x$, the single-mode heat capacity $c_B(z)=z^2/[4\sinh^2(z/2)]$ is strictly decreasing in $z>0$. Indeed, $d\log c_B/dz=2/z-\coth(z/2)<0$. The bosonic entropy $\Phi_B(x;m)$ defined in the Letter obeys
\begin{align}
    \partial_x\Phi_B&=\frac1x\int_0^\infty\frac{dp}{2\pi}\,p\tanh(\pi p)c_B(z)\,,\\
    \partial_m\Phi_B&=-m\int_0^\infty\frac{dp}{2\pi}\frac{p\tanh(\pi p)}{p^2+m^2}c_B(z)\leq0\,.
\end{align}
At fixed $x$, $\partial_x\Phi_B(x;m)\leq\partial_x\Phi_B(x;0)$. Therefore, along the saddle,
\begin{equation}
    \frac{d}{dx}\Phi_B(x;m(x))=\partial_x\Phi_B+m'(x)\partial_m\Phi_B\leq\partial_x\Phi_B(x;0)\,.
\end{equation}
With $C_T=NC_{T,s}+O(N^0)$, this is precisely $\Psi'(x)\leq\Psi_s'(x)$ for $x>1$, with strict inequality for positive interaction. At $\alpha=0$ the free scalar saturates.

\section*{Curvature along the large-\texorpdfstring{$N$}{N} sextic scalar line}

The strict large-$N$ $O(N)$ sextic theory with interaction $g_6(\phi\cdot\phi)^3/(6N^2)$ has a stable conformal line $0\leq\alpha=\sqrt{g_6}/(4\pi)\leq1$. Let $m$ be the hyperbolic dimensionless thermal mass, determined for $x\geq1$ by $m=\alpha\mathcal I(m,x)$ with $\mathcal I$ defined in Eq.~\eqref{SintI}. Near the conformal temperature, set $\delta=x-1>0$. The joint small-$m$, small-$\delta$ expansion on this branch is
\begin{equation}\label{Ssexticgapexp}
    \mathcal I(m,1+\delta)=\frac{\pi}{8}\delta-\frac{\pi}{2}m+o(|\delta|+m)\,.
\end{equation}
Consequently,
\begin{equation}\label{Ssexticmass}
    m=\frac{\pi\alpha}{4(2+\pi\alpha)}\delta+o(\delta)\,.
\end{equation}
To see the resulting change in entropy, the derivative of $\Phi_B$ above gives
\begin{equation}
    -\lim_{m\to0^+}\frac{\partial_m\Phi_B(1;m)}{m}=\int_0^\infty\frac{dp}{2\pi}\,\frac{\tanh(\pi p)}p c_B(2\pi p)=\frac{\pi}{16}\,.
\end{equation}
Using $\Psi=(128/\pi)\Phi_B$ at leading $N$ therefore gives $\Psi_\alpha(x)-\Psi_s(x)=-4m^2+o(\delta^2)$. It follows that
\begin{equation}\label{Ssexticcurvature}
    \boxed{\Psi_\alpha''(1^+)=\frac{19}{15}-\frac{\pi^2\alpha^2}{2(2+\pi\alpha)^2}\,.}
\end{equation}
At the Bardeen--Moshe--Bander endpoint $\alpha=1$,
\begin{equation}
    \Psi_{\rm BMB}''(1^+)=\frac{19}{15}-\frac{\pi^2}{2(2+\pi)^2}=1.0799967\ldots\,.
\end{equation}
The curvature decreases monotonically along the sextic line, despite the presence of a dimension-one scalar bilinear at leading large $N$.

\section*{Next-to-leading flat-space endpoint checks}

To combine published thermal free energies with stress-tensor normalizations, we use $s=c_ST^2$ and $f=-c_ST^3/3$. The next-to-leading thermal coefficients are given in Refs.~\cite{ChubukovSachdevYe1994,DiatlykPopovWang2024}, while $C_T$ is given in Refs.~\cite{Petkou1995,DiabEtAl2016,GiombiTarnopolskyKlebanov2016}. For the scalar $O(N)$ Wilson--Fisher CFT,
\begin{equation}
    \frac{c_S}{C_T}=48.3376+\frac{1.5557}{N}+O(N^{-2})\,.
\end{equation}
For the critical Gross--Neveu and zero-CS-level conformal QED$_3$ theories, let $N_f$ count two-component complex Dirac fermions. The fermionic $C_T$ formulas of Refs.~\cite{DiabEtAl2016,GiombiTarnopolskyKlebanov2016} use $N=2N_f$. Combining the results yields
\begin{equation}
    \left.\frac{c_S}{C_T}\right|_{\rm GN}=45.3165+\frac{0.0755}{N_f}+O(N_f^{-2})\,,\qquad\left.\frac{c_S}{C_T}\right|_{\rm QED_3}=45.3165+\frac{0.8994}{N_f}+O(N_f^{-2})\,.
\end{equation}
These are endpoint checks in the controlled large-flavor regime, not finite-$N_f$ bounds or hyperbolic-response calculations.

\section*{Flat-space field-theory tests below three dimensions}

With $s=c_ST^{d-1}$, scale invariance gives $p=c_ST^d/d$ for the vacuum-subtracted pressure. We compare each theory $X$ with a massless free scalar in the same dimension. The Bose integral gives $p_s/T^d=\Gamma(d/2)\zeta(d)/\pi^{d/2}$ (see, e.g., Ref.~\cite{BenjaminLeeOoguriSimmonsDuffin}). A convenient reference for $C_{T,s}=d\,\Gamma(d/2)^2/[4(d-1)\pi^d]$ is Eq.~(16) of Ref.~\cite{Petkou1995}. Thus
\begin{equation}\label{Sscalarbenchmark}
    B(d)\equiv\left(\frac{c_S}{C_T}\right)_s
    =\frac{4(d-1)\pi^{d/2}\zeta(d)}{\Gamma(d/2)}\,,
    \qquad R_X(d)\equiv\frac{(c_S/C_T)_X}{B(d)}\,.
\end{equation}
In a two-dimensional CFT with $c_L=c_R=c>0$ and the identity as the cylinder ground state, the plane-to-cylinder map gives $c_S=\pi c/3$, while $C_T=c/(2\pi^2)$ in our normalization. Thus $c_S/C_T=2\pi^3/3=B(2)$ independent of $c$. An expansion about $d=2$ therefore tests the sign of the first correction to $R_X=1$.

\paragraph{Gross--Neveu fixed point.} Write $n_\psi=N_f\operatorname{tr}{\bf 1}$ for the number of Dirac flavors times the spinor trace. The free fermion pressure in units of the scalar pressure is $n_\psi(1-2^{1-d})$ (see, e.g., Eqs.~(4.1) and~(4.4) of Ref.~\cite{BenjaminLeeOoguriSimmonsDuffin}), whereas $C_{T,\mathrm F}/C_{T,s}=n_\psi(d-1)/2$~\cite{DiabEtAl2016}. Hence $R_{\mathrm F}(d)=2(1-2^{1-d})/(d-1)$. In $d=2+\varepsilon$ the Gross--Neveu four-fermion coupling at the fixed point is $O(\varepsilon)$ for fixed $n_\psi>2$. The first-order pressure correction is proportional to $\langle(\bar\psi_i\psi_i)^2\rangle_{\rm free}$. At zero chemical potential the massless propagator is odd under $(\omega_n,\boldsymbol p)\mapsto(-\omega_n,-\boldsymbol p)$, so its vacuum-subtracted coincident limit vanishes in every Wick contraction (see Ref.~\cite{BlaizotMendezGalainWschebor2003} for the explicit $d=2$ calculation). Since $C_T/C_{T,\mathrm F}=1+O(\varepsilon^3)$ by Eq.~(4.57) of Ref.~\cite{DiabEtAl2016}, $R_{\rm GN}$ and $R_{\rm F}$ agree through first order in $\varepsilon$. Expanding $R_{\rm F}(2+\varepsilon)$ gives
\begin{equation}\label{SGNbelowthree}
    R_{\rm GN}(2+\varepsilon)=1-(1-\log 2)\varepsilon+O(\varepsilon^2)\,.
\end{equation}

\paragraph{Critical $O(N)$ model at $N=\infty$.} Petkou and Vlachos computed the finite-size free energy on a periodic slab for $2<d<4$~\cite{PetkouVlachos1999}. Setting the slab width equal to $1/T$ gives the thermal pressure. In units of $T$, write $z=m/T$, $E_q=(q^2+z^2)^{1/2}$ and $\int_{\boldsymbol q}=\int d^{d-1}q/(2\pi)^{d-1}$. At criticality, the pressure per component in units of $T^d$ and its stationarity equation are
\begin{align}
    P_d(z)&=\frac{\Gamma(-d/2)}{2(4\pi)^{d/2}}z^d-\int_{\boldsymbol q}\log(1-e^{-E_q})\,,\label{SONflatpressure}\\
    G_d(z)&\equiv-2\frac{\partial P_d}{\partial z^2}
    =\frac{\Gamma(1-d/2)}{(4\pi)^{d/2}}z^{d-2}
    +\int_{\boldsymbol q}\frac{1}{E_q(e^{E_q}-1)}\,,\qquad G_d(z_*)=0\,.\label{SONflatgap}
\end{align}
For $2<d\leq3$, both terms in $G_d(z)$ decrease strictly with $z$. The thermal term diverges at $z=0$ and the vacuum term tends to $-\infty$ as $z\to\infty$. There is a unique positive root $z_*$, and $P_d'(z)=-zG_d(z)<0$ for $0<z<z_*$. Since $P_d(0)=\Gamma(d/2)\zeta(d)/\pi^{d/2}$ is the free-scalar pressure in the same units and $C_T=NC_{T,s}+O(N^0)$~\cite{Petkou1995}, it follows that
\begin{equation}\label{SONflatbound}
    R_{O(\infty)}(d)=\frac{P_d(z_*)}{P_d(0)}<1\,,\qquad 2<d\leq3\,.
\end{equation}
This ratio is $\tilde c/N$ in Fig.~2 of Ref.~\cite{PetkouVlachos1999}. Numerically we obtain $0.90864$ at $d=2.1$ and $0.80406$ at $d=2.5$, recovering the exact $4/5$ at $d=3$~\cite{Sachdev1993}.

To analyze the ratio $R_{O(\infty)}(d)$ near $d=2$, it is useful to re-express Eq.~\eqref{SONflatgap} as a sum over Matsubara frequencies, $G_d(z)=\sum_n\int_{\boldsymbol q}(q^2+\omega_n^2+z^2)^{-1}$, with $\omega_n=2\pi n$. For $2<d<3$, integrating over $\boldsymbol q$ and expanding only the $n\ne0$ terms in $z^2/\omega_n^2$ gives
\begin{equation}
    \frac{G_d(z)}{a_d}=\sum_{n\in\mathbb Z}(\omega_n^2+z^2)^{(d-3)/2}
    =z^{d-3}+2(2\pi)^{d-3}\zeta(3-d)+O(z^2)\,,
\end{equation}
where $a_d=\Gamma((3-d)/2)/(4\pi)^{(d-1)/2}$ and the sum is dimensionally continued.
At $d=2+\varepsilon$, the $z$-independent term is $-1/(\pi\varepsilon)+O(1)$, and balancing it against the zero mode shows $z_*$ is $O(\varepsilon)$~\cite{ChamatiTonchev2000}. Dropping the $O(z^2)$ terms in $G_d(z)$ changes $z_*$ only by $O(\varepsilon^4)$, so $z_*^{d-3}=-2(2\pi)^{d-3}\zeta(3-d)$ to this accuracy. Integrating $P_d'(z)=-zG_d(z)$ to this root then gives $P_d(z_*)-P_d(0)=-\frac{3-d}{2(d-1)}a_dz_*^{d-1}$ to the same accuracy. Dividing by $P_d(0)$ and taking $N\to\infty$ before expanding in $\varepsilon$, we obtain
\begin{equation}\label{SONbelowthree}
    R_{O(\infty)}(2+\varepsilon)=1-\frac32\varepsilon-3\varepsilon^2\log\varepsilon-\kappa\,\varepsilon^2+O(\varepsilon^3\log^2\varepsilon)\,,\qquad
    \kappa=\frac32\left[2\gamma-2+\log\frac\pi2-\frac{\zeta'(2)}{\zeta(2)}\right]=0.26396\ldots\,,
\end{equation}
where $\gamma$ is Euler's constant.
The linear and logarithmic corrections can also be obtained by expanding at fixed finite $N$, as we now discuss.

\paragraph{Nonlinear sigma models.} Consider a sigma model with coordinates $\phi^a$ on an $n_M$-dimensional target $M$ and action $\frac12\int d^dx\,G_{ab}(\phi)\partial_\mu\phi^a\partial^\mu\phi^b$. We consider the one-loop ultraviolet fixed points in $d=2+\varepsilon$ for which the target metric is Einstein, $\mathcal R_{ab}=2\pi\varepsilon G_{ab}$~\cite{Friedan1980,Friedan1985}, so that a complete target is compact~\cite{Myers1941}. Examples are the principal chiral models and the sphere $S^{N-1}$ with $G$ equal to $(N-2)/(2\pi\varepsilon)$ times the unit-sphere metric, which describes the critical $O(N)$ model at finite $N$~\cite{BrezinZinnJustin1976}.

At first order in $\varepsilon$ a single curvature tensor can enter $C_T$, and it contracts only to the scalar curvature $\mathcal R=2\pi\varepsilon n_M$. The ratio $C_T/(n_MC_{T,s})$ is therefore the same for every target at this order. The $O(N)$ result $C_T/[(N-1)C_{T,s}]=1+3\varepsilon^2/[4(N-2)]+O(\varepsilon^3)$~\cite{DiabEtAl2016} then gives $C_T=n_MC_{T,s}[1+O(\varepsilon^2)]$.

In dimensional regularization the two-loop correction to the pressure vanishes, because it contains the coincident-point value of $\partial_\mu\xi^a\partial^\mu\xi^b$ for the fluctuation field $\xi^a$. The modes that vary around the thermal circle therefore give $n_M$ times the free-scalar pressure at this order. The static modes, which are constant around the thermal circle, cannot be treated perturbatively. Their action is $\frac{1}{2T}\int d^{d-1}x\,G_{ab}\partial_i\phi^a\partial_i\phi^b$, which at $d=2$ describes a particle on $M$ with metric $G/T$, the spatial coordinate playing the role of Euclidean time. In dimensional regularization its Hamiltonian is $-\frac12\nabla^2+\frac18\mathcal R[G/T]$, where $\nabla^2$ is the Laplacian of $G/T$ and spheres have $\mathcal R>0$~\cite{FateevOnofriZamolodchikov1993,BastianelliCorradiniNieuwenhuizen2000a,BastianelliCorradiniNieuwenhuizen2000b}. On a compact Einstein target the ground state is the constant wave function, with energy $E_0=\frac18\mathcal R[G/T]=\frac{\pi\varepsilon n_M}{4}T$. The static modes thus lower the pressure by $TE_0$, compared with $n_M\pi T^2/6$ for $n_M$ free scalars in $d=2$. In $D=1+\varepsilon$ spatial dimensions, the static ground-state energy is replaced by a free-energy density. Dimensional analysis gives this density a factor $(\mathcal R/n_M)^{D/(2-D)}$, rather than $\mathcal R/n_M$ at $D=1$, yielding the additional factor $(2\pi\varepsilon)^{2\varepsilon}$, provided the dimensionless coefficient is regular near $D=1$. Hence, for every Einstein target,
\begin{equation}\label{SNLSMbelowthree}
    R_{\rm NLSM}(2+\varepsilon)=1-\frac32\varepsilon-3\varepsilon^2\log\varepsilon+O(\varepsilon^2)\,.
\end{equation}

For $S^{N-1}$ at large $N$, both the linear and logarithmic corrections agree with Eq.~\eqref{SONbelowthree}. The logarithmic correction has the opposite sign to the leading correction, much like the static-mode contribution near four dimensions discussed below. Target-dependent corrections can enter at order $\varepsilon^2$.

\section*{Multiscalar fixed points in \texorpdfstring{$d=4-\varepsilon$}{d=4-epsilon}}

We may also test the flat-space conjecture below four dimensions. The free-scalar value $B(d)$ of Eq.~\eqref{Sscalarbenchmark} equals $2\pi^6/15$ in $d=4$. We ask whether it bounds $c_S/C_T$ for unitary CFTs in $d\geq3$, using general quartic scalar fixed points. Two effects can oppose the conjectured inequality. One is an infrared contribution at $O(\varepsilon^{3/2})$ from the static Matsubara modes, defined below. The other is thermal ordering, already at $O(\varepsilon)$. Our considerations in this section concern only the flat-space coefficient, not the full hyperbolic-space envelope.

For $N$ real scalars, let $V=\frac{16\pi^2}{4!}\lambda_{ijkl}\phi_i\phi_j\phi_k\phi_l$, with real, fully symmetric $\lambda_{ijkl}$. The one-loop fixed-point equation is~\cite{RychkovStergiou2019,OsbornStergiou2018}
\begin{equation}\label{Smultibeta}
    \varepsilon\lambda_{ijkl}=\lambda_{ijmn}\lambda_{mnkl}+\lambda_{ikmn}\lambda_{mnjl}+\lambda_{ilmn}\lambda_{mnjk}\,.
\end{equation}
The leading interaction correction to the flat-space thermal free-energy density is $2\pi^2\lambda_{iijj}(T^2/12)^2$~\cite{DolanJackiw1974}. Relative to $N$ free scalars in the same dimension, the symmetric thermal saddle gives
\begin{equation}\label{SmulticS}
    \frac{c_S}{c_{S,\rm free}}=1-\frac{5}{4N}\lambda_{iijj}+O(\varepsilon^{3/2})\,.
\end{equation}
At one loop the fields acquire a thermal mass matrix $M_{ij}^2=\frac{2\pi^2}{3}T^2B_{ij}$, where $B_{ij}=\lambda_{ijkk}$. Equation~\eqref{SmulticS} presumes that $B$ is positive semidefinite, $B\succeq0$, so that the symmetric saddle is locally stable. Since $C_T/(NC_{T,s})=1+O(\varepsilon^2)$~\cite{OsbornStergiou2018}, the same leading correction applies to $(c_S/C_T)/(c_S/C_T)_{\rm free}$. The trace $\lambda_{iijj}$ is also the sum of the one-loop anomalous dimensions of scalar bilinears~\cite{OsbornStergiou2021,HogervorstToldo2021}.

Contracting Eq.~\eqref{Smultibeta} yields the positive sum
\begin{equation}\label{Smultitrace}
    \varepsilon\lambda_{iijj}=B_{ij}B_{ij}+2\lambda_{ijkl}\lambda_{ijkl}>0
\end{equation}
for every non-Gaussian real fixed point. Moreover, $B_{ij}B_{ij}\geq(\lambda_{iijj})^2/N$ and $\lambda_{ijkl}\lambda_{ijkl}\geq3(\lambda_{iijj})^2/[N(N+2)]$. Thus $\lambda_{iijj}\leq N(N+2)\varepsilon/(N+8)$~\cite{OsbornStergiou2021,HogervorstToldo2021}, saturated by the $O(N)$ Wilson--Fisher fixed point. Consequently,
\begin{equation}\label{Smultibound}
    1-\frac54\frac{N+2}{N+8}\varepsilon\leq\frac{c_S}{c_{S,\rm free}}<1
    \qquad\text{at leading order in }\varepsilon\,.
\end{equation}
The free scalar therefore maximizes the thermal coefficient among symmetric-saddle fixed points to this order, while the $O(N)$ model has the largest leading reduction.

The next correction has the \emph{opposite} sign. It comes from the static Matsubara modes, which are the components of $\phi_i$ that are constant around the thermal circle. These modes are massless at tree level, so $M^2$ is their entire mass. When $B\succeq0$ they contribute $+\frac{5\sqrt6}{3N}\operatorname{tr}B^{3/2}$ to Eq.~\eqref{SmulticS}. For the $O(N)$ model, comparison of the periodic Casimir amplitude in Eq.~(19) of Ref.~\cite{Diehl2006} with the \emph{free} amplitude in the same dimension gives
\begin{equation}\label{SmultiWF}
    \frac{c_S^{O(N)}}{c_{S,\rm free}}=1-\frac54 q\varepsilon
    +\frac{5\sqrt6}{3}(q\varepsilon)^{3/2}+O(\varepsilon^2)\,,
    \qquad q=\frac{N+2}{N+8}\,.
\end{equation}
The dimension-dependent free-field terms in that reference cancel upon taking the ratio. At fixed $N$, the positive $O(\varepsilon^{3/2})$ term cannot overwhelm the negative $O(\varepsilon)$ term for sufficiently small $\varepsilon$. Extrapolation to $\varepsilon=1$ is unreliable: the Ising truncation gives $1-0.42+0.79\simeq1.37$, whereas the three-dimensional Monte Carlo value is $c_S/c_{S,\rm free}\simeq0.800$~\cite{BulgarelliCaselle2025}.

\paragraph{Thermal order.} A negative eigenvalue of $B$ makes the symmetric saddle unstable, and the dimensionally reduced theory on $\mathbb R^{d-1}$ may then enter an ordered phase~\cite{Chai2020,Weinberg1974}. The zero-temperature quartic potential is still nonnegative at a real one-loop fixed point, since Eq.~\eqref{Smultibeta} implies
\begin{equation}
    \varepsilon\lambda_{ijkl}v_iv_jv_kv_l=3(\lambda_{ijmn}v_iv_j)(\lambda_{klmn}v_kv_l)\geq0
\end{equation}
for all real $v_i$~\cite{RychkovStergiou2019}. The leading free energy must be evaluated at the thermal minimum~\cite{Chai2020,DolanJackiw1974}:
\begin{equation}\label{SmultiW}
    \frac{c_S}{c_{S,\rm free}}=1-\frac{5}{4N}\lambda_{iijj}-\frac{30}{N}W+O(\varepsilon^{3/2})\,,\qquad
    W=\min_\phi U(\phi)\leq0\,.
\end{equation}
Here $U(\phi)=B_{ij}\phi_i\phi_j/T^2+2\lambda_{ijkl}\phi_i\phi_j\phi_k\phi_l/T^4$, and $\frac{\pi^2}{3}T^4U(\phi)$ is the leading effective potential of the static modes. Ordering \emph{increases} $c_S$ already at $O(\varepsilon)$, so the argument leading to Eq.~\eqref{Smultibound} does not cover ordered phases.

Equation~\eqref{SmultiW} can be written as
\begin{equation}\label{SmultiQ}
    \frac{c_S}{c_{S,\rm free}}=1-\frac{5}{4N}\lambda_{iijj}(1-Q)+O(\varepsilon^{3/2})\,,\qquad Q=-\frac{24W}{\lambda_{iijj}}\,.
\end{equation}
Here $Q$ is the fraction of the leading interaction correction that ordering cancels. So $c_S$ stays below its free value at $O(\varepsilon)$ if and only if $Q<1$.

As an example, split the $N=n+1$ scalars into a single scalar $\sigma$ and an $O(n)$ vector $\boldsymbol\phi$. The $O(1)\times O(n)$-invariant quartic potential is
\begin{equation}\label{Smultibiconical}
    V(\sigma,\boldsymbol\phi)=2\pi^2\varepsilon[a\sigma^4+b(\boldsymbol\phi^2)^2+2g\sigma^2\boldsymbol\phi^2]\,.
\end{equation}
The couplings $(a,b,g)$ are the $(\alpha,\beta,\gamma)$ of Ref.~\cite{Chai2020} with $m=1$, and Eq.~\eqref{Smultibeta} reduces to their Eqs.~(3.6)--(3.8). The matrix $B$ is diagonal, with entry $\varepsilon b_1$ for $\sigma$ and $\varepsilon b_2$ for each component of $\boldsymbol\phi$, where $b_1=3a+ng$ and $b_2=(n+2)b+g$. For $n\geq18$ the only fixed point with $g\neq0$, apart from the $O(n+1)$ one, is the biconical fixed point. It has $b_1<0<b_2$. So $\sigma$ condenses and breaks the $\mathbb Z_2$ symmetry $\sigma\to-\sigma$ at every temperature. The minimum of $U$ lies at $\boldsymbol\phi=0$ and $\sigma^2=-b_1T^2/(12a)$, so $W=-\varepsilon b_1^2/(24a)$. With $\lambda_{iijj}=\operatorname{tr}B=\varepsilon(b_1+nb_2)$, this gives $Q=b_1^2/[a(b_1+nb_2)]$. The fixed-point equations simplify this to
\begin{equation}\label{SmultiQbiconical}
    1-Q=\frac{\varepsilon[6(n-1)a-8ng]}{\lambda_{iijj}}\,.
\end{equation}
They also give $a(1-9a)=ng^2$, so $a>0$. Then $b_1<0$ requires $g<0$. Since $\lambda_{iijj}>0$ by Eq.~\eqref{Smultitrace}, we have $Q<1$ for every $n$. Numerically, $Q\simeq0.512$ at $n=100$ and $Q\simeq0.787$ at $n=300$. Solving the fixed-point equations for every $n$ from 18 to $10^7$ shows that $Q$ increases with $n$. So ordering cancels a growing fraction of the correction, but never all of it at finite $n$.

At large $n$ the fixed point approaches $(a,b,g)\sim(9,1,-3)/n$, and Eq.~\eqref{Smultibiconical} becomes
\begin{equation}\label{Smultiflat}
    V(\sigma,\boldsymbol\phi)\sim\frac{2\pi^2\varepsilon}{n}(3\sigma^2-\boldsymbol\phi^2)^2\,,
\end{equation}
whose zero-temperature flat direction $3\sigma^2=\boldsymbol\phi^2$, related to the large-rank moduli space of Ref.~\cite{Chai2020}, lets the condensate cancel the leading thermal masses of the $n$ vector fields, in line with $1-Q=78/n+O(n^{-2})$ from Eq.~\eqref{SmultiQbiconical}. The independent large-$n$ analysis of Ref.~\cite[Eq.~(5.8)]{SmolkinYung2026} shows that the thermal coefficient then equals that of $n$ free scalars at leading order throughout $3<d<4$, saturating the $c_S/C_T$ bound although the CFT remains interacting, and that next-to-leading corrections lift the flat direction and select the $\mathbb Z_2$-ordered vacuum found above.

The dimensional reduction also cautions against continuing the ordered saddle to $d=3$. Continuous finite-temperature symmetry breaking is forbidden on flat $\mathbb R^2$, but discrete ordering remains possible. Indeed, Smolkin and Yung find that the $\mathbb Z_2$ ordering of the large-$n$ biconical model persists to arbitrarily high temperature in $d=3$~\cite{SmolkinYung2026}. None of these flat-space results yet establishes an inequality for the response on $S^1\times H^{d-1}$, where entangling-surface terms involving scalar bilinears may also require care.

\section*{Planar Chern--Simons-matter theories}

\paragraph{Common planar endpoint bound.} Combining Eqs.~(7.1), (7.4), (7.6), (7.9), and (7.11) of Aharony \emph{et al.}~\cite{AharonyThermal} with the planar stress-tensor normalization~\cite{AharonyCorrelators} gives, for each of the regular/critical boson and fermion theories,
\begin{equation}
    \frac{c_S}{C_T}=16\pi\,\mathcal G(\mu,\varphi)\,,\qquad\mathcal G(\mu,\varphi)=\frac1{12}\int_\mu^\infty\frac{t(t^2-\mu^2)}{\cosh t-\cos\varphi}\,dt\,,
\end{equation}
where $\mu\geq0$ is the magnitude of the dimensionless thermal mass. For bosons $\varphi=\pi|\lambda|$, while for fermions $\varphi=\pi(1-|\lambda|)$. The integral is our rewriting of the polylogarithmic function in Eq.~(7.1) of Ref.~\cite{AharonyThermal}:
\begin{equation}
    \mathcal G(\mu,\varphi)=\frac{1}{\sin\varphi}\,\mathrm{Im}\left[\frac{\mu^2}{3}\operatorname{Li}_2(z)+\mu\operatorname{Li}_3(z)+\operatorname{Li}_4(z)\right],\quad z=e^{-\mu+i\varphi}\,.
\end{equation}
This follows by inserting $1/(\cosh t-\cos\varphi)=(2/\sin\varphi)\sum_{n\geq1}e^{-nt}\sin(n\varphi)$ and integrating term by term. We assume the standard planar uniform-holonomy thermodynamic saddle. For $0<\varphi<\pi$ the integrand is nonnegative, and the function $\mathcal G$ decreases when either $\mu$ or $\varphi$ is increased:
\begin{align}
    \partial_\mu\mathcal G&=-\frac\mu6\int_\mu^\infty\frac{t\,dt}{\cosh t-\cos\varphi}\leq0\,,\\
    \partial_\varphi\mathcal G&=-\frac{\sin\varphi}{12}\int_\mu^\infty\frac{t(t^2-\mu^2)\,dt}{(\cosh t-\cos\varphi)^2}<0\,.
\end{align}
As $\mathcal G(0,0)=\zeta(3)$, the endpoint bound $c_S/C_T\leq16\pi\zeta(3)$ follows for all four theories, with equality only in the free-scalar limit. This is a flat-space endpoint result, not a calculation of the full $S^1\times H^2$ response. It assumes the standard planar free energies and the appropriate physical thermal saddles; it does not exclude arbitrary off-shell or inhomogeneous phases.

\paragraph{Regular-boson sextic line.} We illustrate the bound over two couplings in the regular-boson (RB) $U(N)$ theory and check the physical saddle using the phase analysis of Dey \emph{et al.}~\cite{Dey2018}. The scalar interaction is
\begin{equation}\label{Saction}
    \delta\mathcal L=\frac{\lambda_6}{6N^2}(\phi^\dagger\phi)^3\,,\qquad \lambda=\frac{N}{k}\,,\quad 0\leq\lambda<1\,,
\end{equation}
where $k$ is the renormalized Chern--Simons level. We use parity to restrict to nonnegative $\lambda$. At strict $N=\infty$ the sextic coupling is marginal. We work at the conformal point with vanishing relevant mass and quartic deformations.

\paragraph{Thermal mass and ungauged limit.} Set
\begin{equation}\label{Shat}
    \hat\lambda=\left(\lambda^2+\frac{\lambda_6}{8\pi^2}\right)^{1/2},\qquad \mu\equiv \frac{M}{T}\,,\qquad z=e^{-\mu+i\pi\lambda}\,.
\end{equation}
Here $M$ denotes the self-consistent thermal mass of the fundamental boson in the unHiggsed saddle on $S^1_\beta\times\mathbb R^2$, with $T=1/\beta$. The positive phase in $z$ matches the convention used in $\mathcal G$ above. For the unHiggsed thermal saddle, Eq.~(7.10) of Ref.~\cite{AharonyThermal} is equivalent to
\begin{equation}\label{Sgap}
    \boxed{\left(1+\frac{2}{\hat\lambda}\right)\mu=\frac{2}{\pi\lambda}\operatorname{Im}\operatorname{Li}_2(z)\,.}
\end{equation}
The $\lambda\to0$ limit is taken continuously. At the conformal point, $\mu$ depends on the couplings but not on temperature; the regular-boson ratio is therefore $16\pi\mathcal G(\mu,\pi\lambda)$ from the common expression above.

The ungauged limit provides a consistency check against the sextic scalar theory discussed earlier. Write $N_r=2N_c$ for the number of real scalar components corresponding to $N_c$ complex scalars. With $\phi^\dagger\phi=\varphi^2/2$, comparison of $g_6(\varphi^2)^3/(6N_r^2)$ with the interaction above gives $\lambda_6=2g_6$. Thus, for $\alpha=\sqrt{g_6}/(4\pi)$, one has $\hat\lambda\to2\alpha$ as $\lambda\to0$. Using $\operatorname{Im}\operatorname{Li}_2(e^{-\mu+i\pi\lambda})=-\pi\lambda\log(1-e^{-\mu})+O(\lambda^3)$, Eq.~\eqref{Sgap} becomes
\begin{equation}\label{Sungauged}
    \left(1+\frac1\alpha\right)\mu=-2\log(1-e^{-\mu})\,.
\end{equation}
It agrees with the $x\to\infty$ limit of the hyperbolic sextic gap equation $m=\alpha\mathcal I(m,x)$, identifying $\mu=2\pi m/x$. In this limit, $\mathcal I(m,x)=-m-(x/\pi)\log(1-e^{-\mu})+o(x)$. At $\alpha=1$ the solution is $\mu=\log 2$, reproducing the Bardeen--Moshe--Bander endpoint.

\paragraph{Stable couplings and phases.} For $0<\lambda<1$ define the sextic parameter used in Ref.~\cite{Dey2018},
\begin{equation}
    x_6=\frac{\lambda_6}{24\pi^2\lambda^2}-1\,.
\end{equation}
Requiring the zero-temperature large-$N$ effective potential to be bounded below gives $\phi_1\leq x_6\leq\phi_2$, with
\begin{equation}\label{Sphi}
    \phi_1=\frac43\left[\frac1{(2-\lambda)^2}-1\right],\qquad\phi_2=\frac43\left[\frac1{\lambda^2}-1\right].
\end{equation}
Equivalently, writing $u\equiv\lambda_6/(8\pi^2)$, the domain is
\begin{equation}\label{Sdomain}
    \boxed{\lambda^2\left[\frac4{(2-\lambda)^2}-1\right]\leq u\leq4-\lambda^2\,.}
\end{equation}
We use the positive-mass unHiggsed solution of Eq.~\eqref{Sgap}; its selection over the Higgsed branch is justified below.

\paragraph{Numerical check and contour map.} Solving Eq.~\eqref{Sgap} and evaluating $16\pi\mathcal G(\mu,\pi\lambda)$ gives at $\lambda=1/2$:
\begin{equation}\label{Sbench}
    \begin{array}{c|c|c|c}
        \lambda_6 & x_6& \mu=M/T & c_S/C_T\\ \hline
        6\pi^2&0&0.298604445&49.0337677\\
        12\pi^2&1&0.342625525&48.8231125
    \end{array}\,.
\end{equation}
These theories have equal planar $C_T$ and separated-point $\langle TTT\rangle$ at fixed $\lambda$, but different thermal coefficients: the sextic interaction changes the thermal mass while not changing the leading spinning-current two- and three-point functions~\cite{AharonyCorrelators}. The entire sampled stable domain lies below the proposed scalar upper bound.

\begin{figure}[ht]
    \centering
    \includegraphics{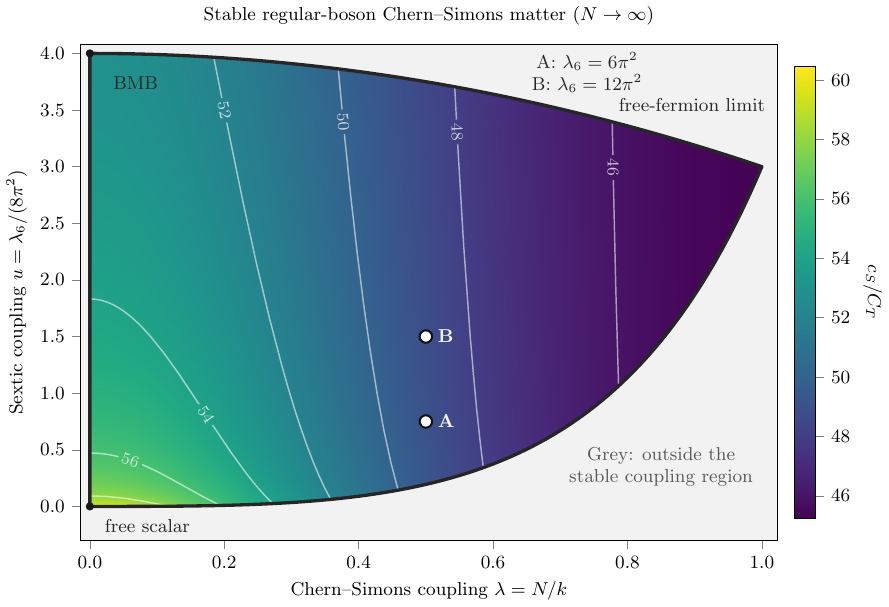}
    \caption{Planar RB Chern--Simons thermal ratio $c_S/C_T$, obtained by solving Eq.~\eqref{Sgap} over the stable domain \eqref{Sdomain}. The axes are $\lambda=N/k$ and $u=\lambda_6/(8\pi^2)$; the black boundaries are the large-$N$ stability limits, and grey denotes couplings outside them. Labelled points A and B are the benchmarks in Eq.~\eqref{Sbench}. The $\lambda=0$ edge contains the free scalar and the Bardeen--Moshe--Bander (BMB) sextic endpoint. The limiting tip $\lambda\to1$ approaches the free-fermion thermal ratio. Values shown are for the planar uniform-holonomy saddle.}
    \label{fig:Scontours}
\end{figure}

The limiting values identify familiar theories. At $(\lambda,\lambda_6)=(0,0)$, the massless free scalar gives $c_S/C_T=16\pi\zeta(3)$; at $(0,32\pi^2)$, the ungauged Bardeen--Moshe--Bander sextic endpoint gives $c_S/C_T=14\pi\zeta(3)$. At the narrowing tip $(\lambda,\lambda_6)\to(1,24\pi^2)$, the ratio approaches the free-Dirac value $12\pi\zeta(3)$.
A plot is shown as Fig.~\ref{fig:Scontours}.

\paragraph{Exclusion of a competing Higgsed saddle.} We use the positive-mass unHiggsed solution of Eq.~\eqref{Sgap} within this domain. To check whether it competes with a Higgsed phase, we specialize the finite-temperature saddle equations of Ref.~\cite{Dey2018}, Sec.~4, to zero mass and quartic couplings. Write $c>0$ for the dimensionless thermal mass in a \emph{putative Higgsed} saddle (this is the $\hat c_B$ of Ref.~\cite{Dey2018}), and denote the normalized Higgs condensate by $\sigma=2\pi v^2/(\lambda T)>0$. The bosonic thermal tadpole, averaged over the gauge holonomy, is
\begin{align}
    \mathcal S(c)&=\frac12\int_{-\pi}^{\pi}d\theta\,\rho(\theta)\left[\log\!\left(2\sinh\frac{c+i\theta}{2}\right)+\log\!\left(2\sinh\frac{c-i\theta}{2}\right)\right],\label{Stadpole}\\
    &=\frac c2-\frac1{\pi\lambda}\operatorname{Im}\operatorname{Li}_2(e^{-c+i\pi\lambda})\,,\nonumber
\end{align}
where $\rho(\theta)=1/(2\pi\lambda)$ for $|\theta|\leq\pi\lambda$ and vanishes otherwise. The second line follows by integrating the logarithms over the uniform holonomy distribution.

Dey \emph{et al.} introduce an off-shell free energy depending on $c$, $\sigma$, and an auxiliary tadpole variable $\widetilde{\mathcal S}$. Its stationarity conditions, Eqs.~(4.3)--(4.5) of Ref.~\cite{Dey2018}, set $\widetilde{\mathcal S}=\mathcal S(c)$ on-shell and separate the two branches through
\begin{equation}\label{Sbranch}
    \bigl(\widetilde{\mathcal S}+\sigma\bigr)\bigl[-c+\lambda(\widetilde{\mathcal S}+\sigma)\bigr]=0\,.
\end{equation}
The first factor describes the unHiggsed branch; on the Higgsed branch the second factor vanishes. The resulting first condition below selects the Higgsed branch [Eq.~(4.5) of Ref.~\cite{Dey2018}], while the second follows from the Higgs-field tadpole condition [Eq.~(3.30) of Ref.~\cite{Dey2018}] at vanishing mass and quartic deformations:
\begin{equation}\label{Shiggs}
    c=\lambda\bigl(\mathcal S(c)+\sigma\bigr)\,,\qquad c^2=\lambda^2(4+3x_6)\sigma^2\,.
\end{equation}
Because the Higgs condensate has $\sigma>0$, these equations imply
\begin{equation}\label{Shiggscondition}
    \frac{\lambda\mathcal S(c)}c=1-\frac1{\sqrt{4+3x_6}}\,.
\end{equation}
For $0<\lambda<1$ and $c>0$, the imaginary part of the dilogarithm in Eq.~\eqref{Stadpole} is positive, so the left-hand side of Eq.~\eqref{Shiggscondition} is strictly smaller than $\lambda/2$. On the other hand $x_6\geq\phi_1$ implies $\sqrt{4+3x_6}\geq2/(2-\lambda)$, hence the right-hand side is at least $\lambda/2$. Thus, no positive-mass Higgsed saddle competes at the conformal point within the stable interval. This statement concerns the standard planar uniform-holonomy construction, not the general finite-$N$ phase diagram.

\end{document}